\documentclass[a4paper,fleqn,usenatbib]{mnras}
\usepackage{mathptmx}

\usepackage[T1]{fontenc}
\usepackage{ae,aecompl}

\usepackage{graphicx}	
\usepackage{amsmath}	
\usepackage{amssymb}	

\title[L2 Leakage - Main Sequence EMRI]{Mass loss through the L2 Lagrange point - Application to Main Sequence EMRI}

\author[I. Linial and R. Sari]{
Itai Linial\thanks{E-mail: itai.linial@mail.huji.ac.il}
and Re'em Sari
\\
Racah Institute of Physics, The Hebrew University, Jerusalem 91904, Israel
}

\date{Accepted XXX. Received YYY; in original form ZZZ}

\pubyear{2017}

\begin{document}
\label{firstpage}
\pagerange{\pageref{firstpage}--\pageref{lastpage}}
\maketitle

\begin{abstract}
We consider stable mass transfer from the secondary to the primary of an extreme mass ratio binary system. We show that when the mass transfer is sufficiently fast, mass leakage occurs through the outer Lagrange point L2, in addition to the usual transfer through L1. We provide an analytical estimate for the mass leakage rate through L2 and find the conditions in which it is comparable to the mass transfer rate through L1. Focusing on a binary system of a main-sequence star and a super-massive black hole, driven by the emission of gravitational radiation, we show that it may sustain stable mass transfer, along with mass loss through L2. If such a mass-transferring system occurs at our Galactic Centre, it produces a gravitational wave signal detectable by future detectors, such as eLISA. The signal evolves according to the star's adiabatic index and cooling time. For low mass stars, the evolution is faster than the Kelvin-Helmholtz cooling rate driving the star out of the main-sequence. In some cases, the frequency and amplitude of the signal may both decrease with time, contrary to the standard chirp of a coalescing binary. Mass loss through L2, when occurs, decreases the evolution timescale of the emitted gravitational wave signal by up to a few tens of per cent. We conclude that L2 mass ejection is a crucial factor in analyzing gravitational waves signals produced by such systems.
\end{abstract}

\begin{keywords}
binaries: general -- stars: mass-loss -- gravitational waves
\end{keywords}



\section{Introduction}

Binary systems are the primary sources of gravitational waves (GW) targeted by detectors such as LIGO, and in the future, eLISA \citep{LIGO_2014,eLISA_2012,eLISA_2013}. The expected GW signal waveform emitted by binary systems is well-understood when the orbit is sufficiently wide and circular. In this regime, the binary can be modelled as two point masses, in a slow inspiral motion. The GW amplitude of a detached binary system increases as the orbit shrinks, making the detection easier at later times. However, when the orbit decreases, the system's evolution is modified due to the increase of tidal effects and the onset of mass transfer. Since these effects become dominant when the gravitational waves emitted by the system are most likely to be detected, it is important to study their influence on the GW signal.

One class of astrophysical GW sources in the eLISA sensitivity range are extreme-mass-ratio-inspirals (EMRIs). These are binary systems of a super massive black hole (SMBH) and a few-solar mass secondary, slowly inspiralling towards the black hole. Since the GW signal of an EMRI probes the motion of the secondary when it is merely a few Schwarzschild radii away from the centre of the black hole, EMRIs could serve as laboratories for accurately testing general relativity at the strong field regime \citep{eLISA_2012,eLISA_2013}.

The secondary component of an EMRI is usually taken to be a compact object. The self gravity of the compact object greatly exceeds the tides from the SMBH, and so these systems do not experience mass transfer prior to the final plunge to the SMBH. It is interesting to consider a main-sequence star as the secondary instead. Most previous studies have neglected this scenario, claiming that a main-sequence star would be tidally disrupted by the SMBH before evolution through GW emission becomes important \citep{Gair_2013}. This is indeed the case when the star approaches the black hole on a very eccentric orbit, with its pericentre well within the tidal radius, producing a tidal disruption event \citep[e.g.][]{Evans_TDE_1989}. However, if the star is orbiting the black hole in a nearly circular orbit just on the tidal radius, it will transfer mass slowly, without being violently disrupted. A few studies have suggested that these systems may indeed form through gravitational encounters between stars in a cluster around the black hole \citep{Freitag_SgrA_2003}. Later works proposed that EMRIs with very low eccentricities can form via tidal separation of binaries by the SMBH \citep{Miller_Freitag_2005}. Unlike compact objects, an inspiralling star will overfill its Roche lobe at some critical separation, and mass transfer will occur. We discuss the influence of stable mass transfer on the evolution of the GW signal associated with these systems.

\cite{Dai_Blandford_2013} considered a system of a main-sequence star orbiting a SMBH in a circular orbit, sustaining stable mass transfer, similar to the systems we investigate. Their work focuses on constraining the mass and spin of the black hole through the X-ray emission associated with the mass accretion.

Mass transfer in binary systems has been studied extensively in various contexts over the past decades \citep[e.g.][and others]{Rappaport_1982, Hjellming_Webbink_87, SPvdH_1997}. Generally, if the primary is sufficiently massive compared to the secondary, and when mass flows from the secondary to the primary, mass transfer is dynamically stable. Mass then flows through the inner Lagrange point, L1, from the donor to the accretor. In this work we show that if a system evolves fast enough, mass ejection from the secondary may occur through the outer Lagrange point, L2, as well. We derive the conditions for mass ejection through L2 for any orbital evolution mechanism, and apply it in the context of a main-sequence star orbiting a SMBH. Recently, Pejcha, Metzger and Tomida \citep{Pejcha_2014,PMT1,PMT2} have studied L2 mass loss, in the context of binary stars in a common envelope phase, and stellar mergers \citep[e.g.][]{Tylenda_2011, Nandez_2014_Merger}. The mass ejection in these systems occurs when the common envelope, engulfing both stars, has expanded all the way to L2. In contrast, in this work ejection occurs due to sufficiently fast orbital evolution, causing the secondary to exceed its Roche lobe all the way to L2.

As an illustration to the L2 ejection mechanism we are suggesting, consider a pitcher or a jug with two spouts, located at different heights. Water is being poured into the pitcher at some uniform rate. When the water level exceeds the height of the bottom spout, it starts to spill out. If water is added to the pitcher slowly enough, the leakage will occur only from the lower spout. However, if water is added at sufficiently high rate, it will leak from both spouts simultaneously, as the water level will rise to some steady state height above the upper spout. In the systems considered in this paper, the pitcher represents the gravitational potential well of the secondary, with the bottom and top spouts being the L1 and L2 Lagrange points, respectively. The rate of water flow is analogous to the secondary's mass loss rate, determined by the orbital evolution timescale. When the system evolves slowly, mass loss occurs only from the L1 point, as is in the scenario usually considered. However, when the system evolves sufficiently fast, mass is lost from L2 as well.

The outline of the paper is as follows. In section \ref{sec:L2 Mass Loss} we derive the conditions for L2 mass loss to occur, under general conditions. In section \ref{sec:MT_stab}, we discuss the stability of a mass transferring binary system, in which the secondary loses mass through both L1 and L2. In section \ref{sec:GW_Evolution} we study the L2 mass loss for a system driven by the emission of gravitational radiation. Section \ref{sec:MS_EMRI} applies the model to a binary system of a super massive black hole and a main-sequence star. Our conclusions are summarized in section \ref{sec:Discussion}.

\section{The L2 Mass loss} \label{sec:L2 Mass Loss}
We consider a binary system of masses $M_1$ and $M_2$, such that $q=\frac{M_2}{M_1}<1$.
We assume that the two masses are orbiting the centre of mass in circular orbits, and that some process slowly decreases the orbital separation over many orbits. We refer to this as the "\textit{external driving"}. Our primary example for an external driving is the emission of gravitational waves, but in general it could also be dynamical friction, binary tidal evolution or other processes.
As the objects approach each other, the tidal forces between the components increase. Under certain conditions stable mass transfer may begin from the secondary to the primary. The secondary overfills its Roche lobe and mass transfer occurs through the first Lagrange point (L1). Here we show that if the external driving is sufficiently fast, the secondary may also eject mass in the direction opposite to the companion, through the second Lagrange point (L2).

We assume \(q \ll 1\), and estimate the mass loss rate through the L2 Lagrange point. 
We consider the secondary component to be a polytrope of index \(n\), i.e., its equation of state is given by \(P=K\rho^{1+1/n}\) where \(K\) is constant, whereas the primary is taken to be a point mass. We use the Roche potential approximation, in which we adopt a non-inertial frame, rotating with the objects about their centre of mass in the orbital frequency. In this rotating frame, a stationary configuration can be described by the Roche potential, which is given by the gravity of the binary components, and the centrifugal potential. The simplifying assumption here is that the gravitational potential of the secondary is truncated at the leading monopole term, neglecting higher orders caused by its tidal deformation. Since the primary is a point mass, its gravitational potential has no higher order terms. This is an adequate approximation, since polytropes are centrally condensed. For instance, the central density of \(n=3\) polytrope is more than 50 times larger than its mean density \citep{Chandrasekhar_1939}. Typical density profiles of the secondary may cause modest deviations from the simplified Roche potential approximation, quantified by \cite{Rappaport_2013}.

In addition we assume that the secondary is co-rotating with the orbit - its sidereal period matches the orbital period. The conditions for maintaining the secondary tidally locked throughout the evolution are examined in appendix \ref{sec:Tidal_Lock}. The synchronization of the primary is irrelevant for our discussion, as it is a point mass.

At some critical orbital separation, the secondary overfills its Roche lobe, and mass transfer ensues. This separation is known as the \textit{Roche limit}. As we later show, mass loss increases as the secondary overfills it Roche lobe to a greater extent.

As mass is transferred from the secondary to the primary, the semi-major axis evolves both due to the external driving and the redistribution of angular momentum in the system. Mass that leaves the secondary through the inner Lagrange point has specific angular momentum that is similar to the specific orbital angular momentum of the binary. If all of this angular momentum is being restored to the orbit, the mass transfer is said to be conservative. The commonly considered process by which angular momentum is restored to the orbit, is the action of tidal torques from the primary. Alternatively, torques between the secondary and an accretion disc that forms around the primary could restore the angular momentum to the orbit. When some fraction of the angular momentum associated with the mass flow does not return to the orbit (in the case of inefficient tidal torques, for example), the flow is said to be non-conservative \citep{SPvdH_1997}. This might also be the case when the angular momentum of the accreted material simply spins up the primary, and does not return to the orbit. As will be discussed in section \ref{sec:MT_stab}, the loss of mass through L2 makes the mass transfer somewhat non-conservative. Note that we are concerned only with conservation of angular momentum and not mass, since $M_2 \ll M_1$, so that the total mass of the system is roughly constant even if mass from $M_2$ is lost.

The response of the secondary's radius to mass loss is another key element in mass-transferring binary systems \citep{Rappaport_1982, Hjellming_Webbink_87}. The mass loss rate is sensitive to the extent of the Roche lobe overfilling. Since the Roche lobe size and the secondary's radius change due to mass loss, the mass transfer rate depends on their mutual evolution.
The combination of these effects determines the stability of mass loss. When mass transfer is stable, the secondary's mass and the orbit evolve together. This is achieved when the Roche lobe size and the secondary's radius match throughout the mass transfer duration. In an unstable mass transfer, the secondary loses all of its mass in a runaway process, while the orbital separation remains roughly constant \citep{Paczynski_1965, Rasio_1994}.

We denote the evolution timescale by \(\tau\), and write:
\begin{equation} \label{eq:Orbital_MT_time}
    a/|\dot{a}|\sim M_2/|\dot{M_2}|\sim\tau \,.
\end{equation}

Due to the coupling between mass transfer and orbital evolution, the system adjusts such that the secondary is always kept at its Roche limit. The secondary slightly overfills its Roche lobe, maintaining the appropriate mass transfer rate. In section \ref{sec:MT_stab} we address the stability of mass transfer and the relation between the mass and orbit evolution timescales.

Given the orbital separation between the components, \(a\), the size of the Roche lobe in the limit \(q \ll 1\) is approximated by
\begin{equation} \label{eq:RL_size}
    R_{L1} \propto a q^{1/3} \,,
\end{equation}
with the constant of proportionality being approximately \(0.462\), \citep{Paczynski_1971,Eggleton_83}.

\begin{figure}
    \includegraphics[width=\columnwidth]{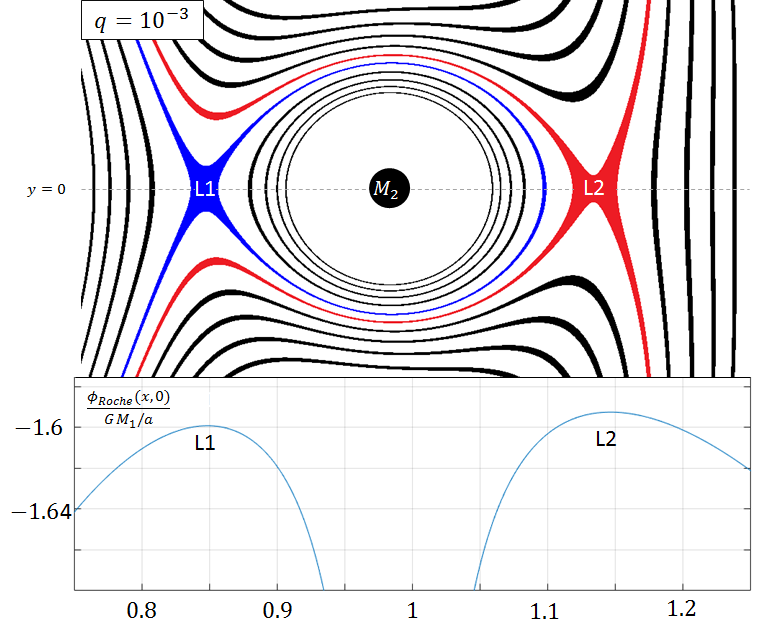}
    \caption{The Roche potential around the secondary, for mass ratio \(q = 10^{-3}\). \textit{Top}: Contours of the Roche potential at the XY plane. Note that the contours have a finite width, since they represent a small potential interval centred around some value. The centre of mass is at the origin \((0,0)\). Blue/Red - the equipotential surface passing through the L1/L2 Lagrange point, respectively. \textit{Bottom}: The potential along the line \(y=0\), as a function of the \(x\) coordinate. Note the potential difference between L1 and L2, which vanishes as \(q\rightarrow0\).}
    \label{fig:Roche_L1_L2}
\end{figure}

Under the Roche approximation, the secondary is not spherical in shape, due to the primary's gravity and the centrifugal potential. The radius of the secondary is thus not well-defined, and we adopt an alternative definition to the secondary's size. Before the Roche lobe is filled, the secondary can attain hydrostatic equilibrium, and its edge follows an equipotential surface (see figure \ref{fig:Roche_L1_L2}). Each equipotential surface, \(\phi\), contained within the Roche lobe has a certain volume, \(V(\phi)\), which corresponds to some \textit{volumetric radius}, \(R(\phi)\), such that \(V(\phi)=\frac{4\pi}{3} R(\phi)^3\). We denote the volumetric radius of the Roche lobe by \(R_{L1}\), since it is the equipotential surface passing through the L1 Lagrange point.

We wish to extend this definition to non-equilibrium configurations, when the secondary overfills its Roche lobe and mass loss occurs. The equipotential surfaces outside the Roche lobe engulf the primary as well, and their volume is much greater than the secondary's actual volume. In slightly out-of-equilibrium configurations, i.e. when the secondary does not exceed the Roche lobe by much, its edge follows an equipotential surface, except at the vicinity of L1 (and possibly L2), where leakage occurs and the secondary does not have clear boundaries. If the secondary extends up to some potential surface \(\phi_2\), where it has a clear boundary, we can approximate the volumetric radius of the secondary by (similarly to the argument used by \cite{Ritter_88})
\begin{equation} \label{eq:secondary_size}
    R_2 \approx R_{L1} + \frac{\phi_2 - \phi_{L1}}{GM_2/R_{L1}^2} \,.
\end{equation}
where \(\phi_{L1}\) is the potential on the Roche lobe. Here we used the fact that on the Roche lobe (except near L1 and L2) the potential gradient is dominated by the secondary's surface gravity, \(|\nabla \phi| \approx GM_2/R_2^2\), and that \(R_2 \approx R_{L1}\), since the secondary just slightly overfills its Roche lobe.

This definition allows us to limit the extent of the leaking secondary by "closing" the appropriate equipotential surface. Since the secondary is considerably deformed by tides when it is at the Roche limit, the value of \(R_2\) deviates from the radius of the secondary had it been isolated, by a factor of order unity \citep{Love_1909}.

We now proceed to discuss the mass transfer rate from the secondary. We derive the mass transfer from the details of the flow at the vicinity of L1. We consider a (semi-) steady-state flow through L1, which changes on the orbital evolution timescale \(\tau\). The mass loss rate of the secondary is given by \(\dot{M_2}=-\rho v Q\). Here \(\rho\) is the ejected material density, \(v\) is its flow velocity, and \(Q\) is the flow cross section (the area of the L1 nozzle). We define the Roche lobe overfilling length as \(\delta \equiv R_2 - R_{L1}\). Similar calculations were first carried out by \cite{Paczynski_Sienkiewicz_72}, and later by \cite{Savonije_1978, Ritter_88, Kolb_Ritter_1990, Ge_2010, Pavlovskii_2015} and many others. We follow similar arguments in order to estimate the value of $\delta$, and then proceed to discuss the L2 mass loss, which was not calculated by these previous works.

The hydrodynamic conditions at the L1 nozzle are given by the conditions at depth \(\delta\) within the secondary. Since we assume that the secondary is a polytrope of index \(n\), the density near the edge is
\begin{equation} \label{eq:PolytropeDensity}
    \rho(\delta)\approx\frac{M_2}{R_2^3} \left( \frac{\delta}{R_2} \right)^n \,,
\end{equation}
 and \(v\approx c_s(\delta)\approx v_{esc} (\delta/R_2)^{1/2}\), where \(c_s(x)\) is the speed of sound at depth \(x\), and \(v_{esc}\approx c_s(R_2)\) is the escape velocity from the secondary's surface, had it been isolated. In other words, the ejected material develops velocities comparable to its speed of sound when at L1 \citep[e.g.][]{Lubow_Shu_1975,Lubow_Shu_1976,Lubow_Shu_1979}. The nozzle's cross-section can be described as a truncated sphere - removing a cap at depth \(\delta\) leaves a nozzle with an area \(Q\approx \delta \cdot R_2\).

Combining all these terms we get:
\begin{equation}
    |\dot{M_2}|/M_2\approx\tau_{dyn}^{-1}(\delta/R_2)^{n+3/2}
\end{equation} where \(\tau_{dyn}\approx R_2/v_{esc}\) is the secondary's dynamical time (or its sound crossing time).
The Roche lobe overfilling extent, \(\delta\), can be determined from equation \ref{eq:Orbital_MT_time}. We obtain:
\begin{equation} \label{eq:delta_tau}
    \delta/R_2\approx(\tau_{dyn}/\tau)^\frac{1}{n+3/2}  
\end{equation}
where we have used equation \ref{eq:Orbital_MT_time}. As expected, faster evolution demands faster mass transfer which in turn, requires larger values of \(\delta\).

The orbital period is approximately \(\sqrt{a^3/GM_1}\), and using equation \ref{eq:RL_size}, we find that at the Roche limit, the orbital period is similar to the dynamical time of the secondary, \(\tau_{dyn}\). The ratio \(\tau / \tau_{dyn}\) which appears in equation \ref{eq:delta_tau} is therefore roughly the number of orbits completed during the system's evolution time.

We note that if \(\delta\) is sufficiently large, matter might leak from the L2 Lagrange point, in addition to the mass transfer to the primary via the L1 point. In order to quantify this suggested effect, we examine the potential difference between these two Lagrange points. Through this potential difference, we estimate the radial gap between the equipotential surfaces of L1 and L2, and compare it to \(\delta\).

The discussed effect of L2 leakage arises from the difference in Roche potential between the L1 and L2 points. The potential at L2 is always higher than the L1 potential, and it is this asymmetry that leads to the qualitative difference between the L1 and L2 flows. We demonstrate that the mass transfer rate through L1 always exceeds the L2 mass transfer rate.

The volumetric radius difference between the two equipotential surfaces can be approximated by
\begin{equation}
    \Delta R_{L_1L_2} \approx \Delta \phi_{L_1L_2} (G M_2/R_2^2)^{-1}  \,,
\end{equation}
where \(\Delta \phi_{L_1L_2}\) is the potential difference between the two surfaces. Here we repeated the procedure used for defining the secondary's size (equation \ref{eq:secondary_size} and subsequent discussion).

For \(q\ll 1\), \(\Delta \phi_{L_1L_2} = \frac{2}{3} q (GM_1/a)\), where \(a\) is the orbital separation \citep{Murray_Dermott_Solar_System_Dynamics}. As \(q\rightarrow0\), the potential difference vanishes, and we get a symmetric mass loss from both Lagrange points. This potential difference is illustrated in figure \ref{fig:Roche_L1_L2}. Put together, we find that the gap between the L1 and L2 surfaces is
\begin{equation} \label{eq:DeltaRL1RL2}
    \frac{\Delta R_{L_1L_2}}{R_2} \approx \frac{\Delta \phi_{L_1L_2}}{GM_2/R_2} \approx \frac{2}{3} \frac{R_2}{a} \approx q^{1/3} \,.
\end{equation}

Coincidentally, this ratio is similar to $R_2/a$, the ratio between the secondary's size and the orbital separation of the binary (equation \ref{eq:RL_size}). We continue our derivation assuming that the L2 mass loss does not impede the mass transfer stability. This assumption is examined in section \ref{sec:MT_stab}, and is found to be generally valid for the relevant parameter range.

Similar arguments regarding the L1 flow apply to \(\dot{M}_{L_2}\), the L2 mass transfer. Therefore, for \(\delta \geq \Delta R_{L_1L_2}\),
\begin{equation}
|\dot{M}_{L_2}|/M_2\approx\tau_{dyn}^{-1}\left(\frac{\delta-\Delta R_{L_1L_2}}{R_2}\right)^{n+3/2} \,.
\end{equation}

The ratio between the mass ejected through L2 to the mass transferred through L1 is thus given by
\begin{equation} \label{eq:ML2_ML1_gen}
|\dot{M}_{L_2}/\dot{M}_{L_1}| = \left(1-\frac{\Delta R_{L_1L_2}}{\delta}\right)^{n+3/2} \,,
\end{equation}
where the above relation holds for \(\delta\geq\Delta R_{L_1L_2}\), i.e. if mass transfer through L2 occurs. As anticipated, the L2 mass loss is always smaller than the L1 mass transfer rate. This has allowed us to estimate the value of $\delta$ by considering the mass transfer rate through the L1 nozzle alone. The total mass loss rate $\dot{M}_2=\dot{M}_{L_1}+\dot{M}_{L_2}$, varies between $\dot{M}_{L_1}$ when no L2 leakage occurs, to $2\dot{M}_{L_1}$, when the leakage is symmetric.

Using equation \ref{eq:delta_tau} and \ref{eq:DeltaRL1RL2}, we obtain the condition for mass flow through L2 as
\begin{equation} \label{eq:LeakageCriterion}
\Delta R_{L_1L_2}/\delta \approx \ q^{1/3} (\tau/\tau_{dyn})^\frac{1}{n+3/2}<1 \,.
\end{equation}

From equation \ref{eq:ML2_ML1_gen} we see that as \(\Delta R_{L_1L_2}/\delta \rightarrow 0\), we get \(\dot{M}_{L_2} = \dot{M}_{L_1}\), implying that \textit{half} of the secondary's mass is ejected through L2, while the other half flows through L1. Unless the parameters are finely tuned such that \(\Delta R_{L_1L_2} \approx \delta\), we expect that either no mass will leak through L2, or that approximately half the mass will. However, as we shall see in section \ref{subsec:L2_MT}, in practice, this transition is rather gradual. Note that here we assumed that the secondary has a sharp boundary, as in \cite{Paczynski_Sienkiewicz_72}, rather than adopting the ''isothermal" model used by \cite{Ritter_88}. In appendix \ref{appendix:ScaleHeight} we show the conditions for L2 mass loss due to large atmospheric scale height, rather than fast system evolution. The relevance of either approximation in the context of L1 mass transfer was discussed by \cite{MNS_DWD_MT} as well.

\section{Mass transfer stability with L2 leakage} \label{sec:MT_stab}

Mass transfer stability is affected by leakage through L2. \cite{Pavlovskii_2015} proposed a new stability criterion for mass transfer in binary systems, depending on whether the donor extends all the way to the outer Lagrange point. In section \ref{sec:L2 Mass Loss} we derived the conditions for which their criterion is met in the \(q \ll 1\) limit, see equation \ref{eq:LeakageCriterion}. In this section we show that the onset of mass transfer from L2 does not necessarily lead to instability. This suggests that as they have stated, their criterion is generally too strict.

Similar discussions appear in \cite{Rappaport_1982}, and in other contexts where non-conservative components of the mass transfer are considered \citep[e.g.][]{SPvdH_1997, Jia_Spruit_2017}.

We assume that the secondary transfers mass at a rate \(\dot{M}_{L_1}=(1-\alpha)\dot{M_2}\) through L1 to the primary, and \(\dot{M}_{L_2}=\alpha \dot{M_2}\) through L2. From equation \ref{eq:ML2_ML1_gen} we have that the mass lost through L2 is always smaller than the mass ejected through L1, so that \(0 \leq \alpha \leq \frac{1}{2}\). When keeping track of the angular momentum budget, we ignore the spin angular momentum of the components. We also assume that as a result of mass loss, the secondary's radius readjusts as follows:
\begin{equation} \label{eq:MassRadius}
    R_2 \propto M_2^{\varepsilon}.
\end{equation}
Note that \(\dot{M_2}<0\), since the secondary is the mass losing component. As before, we take the limit of \(q \ll 1\).

The system is driven by loss of angular momentum through some external driving - gravitational radiation, for instance. We denote the torque associated with this process by \(\dot{J}_{ext} < 0\). We assume that this external driving is slow compared to the orbital period or dynamical timescale of the secondary. For the mass transfer to be stable, the secondary should be at its Roche limit throughout the evolution, slightly overfilling the Roche lobe to sustain the necessary mass transfer rate. 

In tidally locked stellar binaries, angular momentum carried by the mass which flows through L1 to the primary, must eventually find its way back to the orbit. In case where the primary is a black hole, tides are too week to carry angular momentum from the black hole to the orbit on a timescale shorter than the system evolution under gravitational waves \citep{Bildsten_Cutler_1992}. Yet, if the matter is slowly accreted onto the black hole, it must first lose its angular momentum, which presumably ends up in the orbit. This could be manifested by torques between an accretion disc and the secondary's orbit. Therefore, we assume that the orbital angular momentum is unaffected by the L1 mass transfer.
To the contrary, mass transferred through L2 is removed from the system, along with its angular momentum. 
We neglect the torques that the L2 mass applies on the secondary as it leaves.

Therefore, the orbital angular momentum evolves as
\begin{equation}
    \dot{J}_{orb} = \dot{J}_{ext} + \dot{J}_{L_2} \,,
\end{equation}
where \(\dot{J}_{ext}\) is the external driving torque, and
\begin{equation}
    \dot{J}_{L_2} \approx \sqrt{GM_1 a} (\alpha \dot{M_2}) \, 
\end{equation}
is the angular momentum lost through L2. Here we neglected the fact that matter at the L2 point has slightly higher specific angular momentum that the specific orbital angular momentum.

Using \(J_{orb} \approx \sqrt{G M_1 a} M_2\), we obtain
\begin{equation} \label{eq:stab_mt_ang_mom_conserv}
    \frac{\dot{J}_{orb}}{J_{orb}} =   \frac{1}{2}\frac{\dot{a}}{a} + \frac{\dot{M_2}}{M_2} = \frac{\dot{J}_{ext}}{J_{orb}} + \alpha \frac{\dot{M_2}}{M_2} \,,
\end{equation}
where we neglected the relative change in the primary's mass \(\dot{M}_1/M_1\) as it is much smaller than the relative change in the secondary's mass, \(\dot{M}_2/M_2\).

Since the secondary is kept at its Roche limit, equation \ref{eq:RL_size} gives
\begin{equation} \label{eq:stab_mt_adot_mdot}
    \frac{\dot{a}}{a} \approx \left( \varepsilon - \frac{1}{3} \right) \frac{\dot{M_2}}{M_2} \,,
\end{equation}
where we used the mass-radius relation from equation \ref{eq:MassRadius}. 
If the secondary's radius does not shrink too much (\(\varepsilon<\frac{1}{3}\)), the orbit expands
to adjust the size of the Roche lobe to the volume of the secondary. Otherwise, the orbit shrinks over time.

We thus relate the mass transfer rate to the external angular momentum loss using equations \ref{eq:stab_mt_ang_mom_conserv} and \ref{eq:stab_mt_adot_mdot} through
\begin{equation} \label{eq:Mdot_Jdot}
    \left(\frac{5}{6} + \frac{\varepsilon}{2} - \alpha\right) \frac{\dot{M_2}}{M_2} = \frac{\dot{J}_{ext}}{J_{orb}} \,.
\end{equation}

For our assumptions to be consistent, we require that both $\dot{J}_{ext}/J_{orb}$ and $\dot{M}_2/M_2$ are negative - the loss of angular momentum leads to mass loss from the secondary, and so:
\begin{equation}
    \frac{5}{6} + \frac{\varepsilon}{2} - \alpha > 0 \,.
\end{equation}

Since \(\alpha \leq \frac{1}{2}\), we have that the above stability condition is satisfied for \(\varepsilon > -\frac{2}{3}\). We find that the mass lost through L2 tends to destabilize the process, however, for reasonable values of \(\varepsilon\), mass transfer is stable for \(q \ll 1\).

Often in the literature, $\alpha$ parametrises the fraction of angular momentum that is not conserved in the orbit. Here we associate $\alpha$ just with the L2 mass loss, and neglect any other channels of angular momentum loss, such as spinning up the primary.

\section{Evolution through gravitational radiation} \label{sec:GW_Evolution}

We proceed with the analysis of the L2 leakage, when the system is driven by the emission of gravitational waves. The point mass approximation gives the following scaling relation for the evolution timescale due to the gravitational quadrupole radiation \citep{L&L_75}:
\begin{equation} \label{eq:GW_timescale_gen}
\frac{a}{|\dot{a}|}\sim\tau_{GW} = \frac{5}{32} \frac{c^5 a^4}{G^3 (M_1+M_2) M_1 M_2 } \,.
\end{equation}
where we define \(\tau_{GW} \equiv -(J/\dot{J}_{GW}) > 0\), the timescale of angular momentum loss due to GW emission.

For \(q\ll1\), and at the Roche limit, where \(a = a_{RL} \approx R_2 q^{-1/3}\), we can approximate:
\begin{equation} \label{eq:GW_timecale}
    \tau_{GW}\approx\frac{c^5 R_2^4 q^{-4/3}}{G^3 M_2^3 q^{-2} }\approx\frac{c^5 R_2^4}{G^3 M_2^3}q^{2/3} \,.
\end{equation}

Note that \(v_{esc}^2\approx GM_2/R_2\), and so
\begin{equation} \label{eq:GW_timescale_beta}
    \tau_{GW}\approx \tau_{dyn} \beta^{-5} q^{2/3} \,,
\end{equation}
where \(\beta\equiv v_{esc}/c\), quantifies how relativistic the secondary is.

Substituting in equation \ref{eq:LeakageCriterion}, we get:
\begin{equation} \label{eq:depth_ratio_gw}
  \Delta R_{L_1L_2}/\delta \approx \ q^{1/3} (\beta^{-5}q^{2/3})^\frac{1}{n+3/2} \,.
\end{equation}

By demanding that this ratio is smaller than unity, we find that:
\begin{equation}
    q<\beta^{15/(n+7/2)} \,.
\end{equation}

For reasonable values of \(n\), we get quite strong dependence on \(\beta\). For a stellar mass secondary, we take the primary to be a black hole. We now introduce another constrain - the minimal separation between the components, \(a_{min}\) is of the order of the primary's Schwarzschild radius:
\begin{equation}
    a_{min}\approx GM_1/c^2 < a_{RL} \approx R_2 q^{-1/3} \,,
\end{equation}
rearranging, we get that
\begin{equation}
    q>\beta^3 \,.
\end{equation}

Put together, with gravitational radiation evolution, leakage through the L2 point is possible when the following condition is satisfied:
\begin{equation} \label{eq:GW_LeakageCriterion}
\beta^3 < q < \beta^{15/(n+7/2)} \,,
\end{equation} 
where the left inequality is the criterion for the secondary to overfill its Roche lobe outside the Schwarzschild radius of the primary, while the right inequality is the criterion for gravitational radiation to drive the system fast enough to enable L2 leakage.

We note that this set of inequalities could only be satisfied for high enough polytropic index \(n>1.5\).
High \(n\) represents secondaries in which a given density, well below the average density, is obtained relatively deep inside the star, see equation \ref{eq:PolytropeDensity}. For a given mass loss rate, set by the rate of gravitational radiation, this causes the leakage from L1 to occur at high \(\delta\), therefore enabling leakage from L2. Yet, values of \(n>1.5\) are common in main sequence star envelopes, as we see below.

\section{Mass transfer between a main-sequence star and a super-massive black hole} \label{sec:MS_EMRI}

We now consider a binary system consisting of a SMBH and a main-sequence star. The star orbits the SMBH in a slowly decreasing circular orbit by radiating gravitational waves, until reaching its Roche limit. For typical stellar masses and SMBH masses, such systems will emit a strong gravitational wave signal, which could be measured by future GW detectors. When the star has reached its Roche limit, the system continues to evolve through emission of gravitational waves, along with stable mass transfer from the star to the SMBH. As we show below, the presence of this stable mass transfer leaves an imprint on the GW signal emitted from such a system, which depends on the mass-radius relation of the evolving star, as well as the amount of mass leakage through L2. Depending on the masses of the components, the signal may be characterized by a "reverse-chirp" - the GW frequency and strain amplitude decrease with time, in contrast with the common chirp signal of a binary merger. Some main sequence SMBH systems may sustain mass loss through L2, along with the stable L1 mass transfer. When substantial L2 mass loss occurs, the evolution timescale of the system may decrease by up to a few tens of per cent.

\subsection{Characteristic sizes and mass transfer rate}

Mass transfer ensues when the stellar component is driven sufficiently close to the black hole, and its Roche lobe is filled. Since the secondary is a main-sequence star, its radius can be estimated by
\begin{equation} \label{eq:MassRadiusMS}
    R_2^{MS} \approx R_{\odot} \left(\frac{M_2}{M_{\odot}}\right)^{\varepsilon_{MS}} \,,
\end{equation}
where \(\varepsilon_{MS} \approx 0.8\) \citep{Kippenhahn_1994}. Using equation \ref{eq:RL_size}, the orbital separation at the Roche limit is given by
\begin{equation} \label{eq:a_MS_SMBH}
    a \approx 1.5 \left(\frac{M_2}{M_{\odot}}\right)^{0.47} \left(\frac{M_1}{M_{SgrA*}}\right)^{1/3} \; \rm AU \,,
\end{equation}
where the black hole mass was scaled by the mass of the black hole at our Galactic Centre, Sagittarius A*, \(M_{SgrA*}\approx 4\times10^6 M_{\odot}\) \citep{Boehle_2016_SgrA, Gillessen_Genzel_2016}.

The gravitational waves timescale, given by equation \ref{eq:GW_timescale_gen} is
\begin{equation} \label{eq:tau_GW_MS_SMBH}
    \tau_{GW} \approx 3\times10^5 \left(\frac{M_2}{M_{\odot}}\right)^{0.87} \left(\frac{M_1}{M_{SgrA*}}\right)^{-2/3} \rm yr \,.
\end{equation}

The mass transfer rate is therefore:
\begin{equation}
    |\dot{M}_2| \approx M_2/\tau_{GW} \approx 10^{-6} \left(\frac{M_2}{M_{\odot}}\right)^{0.13} \left(\frac{M_1}{M_{SgrA*}}\right)^{2/3} M_{\odot} \; \rm yr^{-1} \,.
\end{equation}

Most, if not all, of this mass transfer passes through L1 and towards the black hole. This inflow rate is an order of magnitude smaller than the Bondi accretion associated with our SMBH, \(\dot{M}_b \approx 10^{-5} M_{\odot} \; {\rm yr^{-1}}\) \citep[e.g.][]{Roberts_faint_GC}, and is therefore unlikely to have a significant effect on the electromagnetic luminosity of the vicinity of the SMBH (which is unexpectedly low in our galaxy). 

\subsection{The GW properties}

The mass transferring systems we discuss evolve through the emission of gravitational radiation. Here we discuss the properties of their GW emission. If the system is located at a distance \(d\) from the observer, the measured GW strain amplitude is approximately \citep{BHWDNS}
\begin{equation} \label{eq:h_GW_gen}
    h \approx \frac{GM_1/c^2}{d} \frac{G M_2/c^2}{a} \approx \frac{1}{d} \frac{GM_1}{c^2} q^{1/3} \beta^2 \,,
\end{equation}
where we used the fact that the star is at its Roche limit.

Using the mass-radius relation of main-sequence stars (equation \ref{eq:MassRadiusMS}), \(\beta_{MS}\propto \sqrt{M_{MS}/R_{MS}}\) is given by
\begin{equation} \label{eq:MS_beta}
    \beta_{MS} = \beta_{\odot} \left(\frac{M_2}{M_{\odot}}\right)^{(1-\varepsilon_{MS})/2} \,,
\end{equation}
where \(\beta_{\odot} = \sqrt{GM_{\odot}/R_{\odot}}/c \approx 1.5\times10^{-3}\), and \((1-\varepsilon_{MS})/2\approx0.1\). The instantaneous strain amplitude is therefore
\begin{equation} \label{eq:h_num}
    h \approx 10^{-19} \left(\frac{M_2}{M_{\odot}}\right)^{0.53} \left(\frac{M_1}{M_{SgrA*}}\right)^{2/3} \left(\frac{d}{8 \; \rm kPc}\right)^{-1} \,,
\end{equation}
where \(d\) is scaled by the distance to Sgr A*, roughly \(8 \; \rm kPc\) away \citep{Boehle_2016_SgrA, Gillessen_Genzel_2016}.

The gravitational waves emitted by this system would have a typical frequency \(2 f_{orb}\) \citep[e.g.][]{Peters_Mathews_1963}, where \(f_{orb}\) is approximated by the Keplerian orbital frequency,
\begin{equation} \label{eq:f_GW_gen}
    f = 2 f_{orb} = \frac{1}{\pi} \sqrt{\frac{G(M_1+M_2)}{a^3}} \approx \frac{1}{\pi} \sqrt{\frac{GM_1}{a^3}} \,.
\end{equation}

Using equation \ref{eq:RL_size}, equation \ref{eq:MassRadiusMS} and plugging in the numerical values, we find the frequency, independently of the black hole's mass is
\begin{equation} \label{eq:f_GW_EMRI}
    f \approx 7\times10^{-5} \left(\frac{M_2}{M_{\odot}}\right)^{-0.7} \rm Hz \,.
\end{equation}

Note that this frequency is similar to the dynamical frequency of the star. The frequency and strain amplitude evolve on the gravitational wave timescale, \(\tau_{GW}\), which is much larger than the measurement duration, \(T\), expected to be of order years. The characteristic strain amplitude of the signal is therefore
\begin{equation}
    h_c \approx h \sqrt{Tf} \,.
\end{equation}

Using equations \ref{eq:h_num} and \ref{eq:f_GW_EMRI}, 
\begin{equation}
    h_c \approx 10^{-17} \left(\frac{M_2}{M_{\odot}}\right)^{0.18} \left(\frac{M_1}{M_{SgrA*}}\right)^{2/3} \left(\frac{d}{8 \; \rm kPc}\right)^{-1} \left(\frac{T}{1 \; \rm yr}\right)^{1/2} \,.
\end{equation}

Since the GW frequency depends only  on the secondary's mass (equation \ref{eq:f_GW_EMRI}), we can write the characteristic amplitude as follows:
\begin{equation}
    h_c \approx 10^{-17} \left(\frac{f}{10^{-4} Hz}\right)^{-0.26} \left(\frac{M_1}{M_{SgrA*}}\right)^{2/3} \left(\frac{d}{8 \; \rm kPc}\right)^{-1} \left(\frac{T}{1 \; \rm yr}\right)^{1/2} \,.
\end{equation}

This result could be directly compared to expected sensitivity curves of various instruments, such as eLISA \citep{eLISA_2012,eLISA_2013}. For EMRIs in our Galactic Centre this emission is above the expected detector sensitivity, yielding a signal-to-noise ratio of up to 10.

\subsection{The mass-radius relation} \label{subsec:MassRadius_EMRI}

We now discuss the value of \(\varepsilon\) - the exponent of the mass-radius relation (see equation \ref{eq:MassRadius}). When stable mass transfer takes place, the secondary remains at the Roche limit, such that $R_2 \approx R_{L1}$. The orbital evolution thus depends on $\varepsilon$, determining whether mass transfer increases or decreases the binary separation, and at what rate. The secondary could evolve according to one of a few different regimes, each corresponds to a different value of \(\varepsilon\). The different evolutionary regimes differ by the GW evolution timescale of the binary compared to the cooling time of the secondary. In this section we use simple stellar models in order to estimate $\varepsilon$ in these various regimes.

The secondary's cooling time (or Kelvin-Helmholtz time, \(\tau_{KH}\)) is given by
\begin{equation} \label{eq:T_KH}
    \tau_{KH} = \frac{GM_2^2}{R_2 L}\,,
\end{equation}
which is the time to radiate the gravitational energy of the star. Here $L$ is the luminosity of the secondary. We shall limit the discussion to stars in which the gas pressure dominates over the radiation pressure. In addition, we will consider stars with internal radiative structure, so that the generated energy propagates by radiation diffusion. These conditions are satisfied for a wide range of main-sequence stellar masses, $0.5 M_\odot \lesssim M_2 \lesssim 25 M_\odot$. Under these assumptions, the luminosity $L$ depends on the opacity \(\kappa\) and internal temperature of the star, \(T\) as
\begin{equation}
    L \approx ( \kappa M_2 / R_2^2)^{-1} 4\pi R^2 \sigma T^4 \,,
\end{equation}
where \(\sigma\) is the Stefan-Boltzmann constant. Equipartition gives \(k_B T\approx GM_2 m_p/R_2\) where \(k_B\) is the Boltzmann constant, and \(m_p\) is the proton mass. Therefore, the luminosity scales as \(L \propto M^3/\kappa\). Depending on the temperature and density in the star, \(\kappa\) is either taken to be constant (Thompson opacity, \(\kappa_{es}\)), or for lower temperatures (and higher densities) it is given by the Kramer's opacity (\(\kappa \propto (M_2/R_2^3) T^{-7/2}\)) \citep{Kippenhahn_1994}.

In the general case, the Kelvin-Helmholtz timescale scales as
\begin{equation} \label{eq:tau_KH_kappa}
    \tau_{KH} \propto \kappa (M_2 R_2)^{-1} \,.
\end{equation}

In the next subsections we consider the various evolutionary regimes.

\subsubsection{Main-sequence evolution - $\tau_{GW}>\tau_{KH}$}

Mass is removed from the secondary on a timescale of order \(\tau_{GW}\) (equation \ref{eq:Orbital_MT_time}). If \(\tau_{GW}\) is longer than the star's cooling time, \(\tau_{KH}\), the star has sufficient time to radiate its excess energy as it loses mass, and it follows the common main-sequence mass-radius relation as in equation \ref{eq:MassRadiusMS}, with \(\varepsilon_{MS}\approx 0.8\). Therefore, the mass losing secondary simply evolves as a main-sequence star, with ever decreasing mass.

Comparing \(\tau_{KH}\) to \(\tau_{GW}\) from equation \ref{eq:GW_timecale}, and scaling by the values of the Sun, we find that for main sequence stars, assuming constant $\kappa$
\begin{multline} \label{eq:timescales_ratio_MS}
    \left. \frac{\tau_{KH}}{\tau_{GW}} \right|_{MS} \approx \frac{G^4 M_{\odot}^5}{c^5 R_{\odot}^5 L_{\odot}} \left(\frac{M_2}{M_{\odot}}\right)^{\frac{13}{3}-5\varepsilon_{MS}-3} \left(\frac{M_1}{M_{\odot}}\right)^{2/3} \\
    \approx 100 \left(\frac{M_2}{M_{\odot}}\right)^{4/3 - 5\varepsilon_{MS}} \left(\frac{M_1}{M_{Sgr A*}}\right)^{2/3} \,.
\end{multline}

Here we note that since the power \(4/3-5\varepsilon_{MS}\) is negative, the above ratio \textit{increases} as \(M_2\) decreases. Starting with a sufficiently massive star, the above ratio is smaller than unity, and $\tau_{GW}>\tau_{KH}$. The star then loses mass until \(M_2=M_{\star}\), where the two timescales become similar \(\tau_{KH}=\tau_{GW}\). This threshold mass is approximately
\begin{equation} \label{eq:M_star}
    M_{\star} = 5.6 M_\odot \left(\frac{M_1}{M_{Sgr A*}}\right)^{0.25}\,.
\end{equation}

Therefore, the system evolves in this regime as long as the secondary is sufficiently massive, \(M_2>M_{\star}\).

\subsubsection{Adiabatic response - $\tau_{GW}<\tau_{KH}$}

If the secondary's initial mass, prior to the onset of mass transfer is smaller than \(M_{\star}\), the Kelvin-Helmholtz time is initially \textit{longer} than the GW timescale. This implies that the response of the star to mass loss is adiabatic. If the stellar material has an effective adiabatic index \(\gamma\), its radius changes as
\begin{equation} \label{eq:adiabatic_radius_mass}
    R_2^{ad}\propto M_2^{(2-\gamma)/(4-3\gamma)} \,.
\end{equation}

We denote \(\varepsilon_{ad} = (2-\gamma)/(4-3\gamma)\), which is negative for typical values of \(\gamma\). For low mass stars, where radiation pressure is negligible, $\gamma=5/3$ resulting in \(\varepsilon_{ad} = -1/3 \). For more massive stars, where radiation pressure becomes more significant, the adiabatic index approaches $4/3$ and $\varepsilon_{ad}$ becomes a large negative number. Since the initial stellar mass at this regime is smaller than $M_\star$, substantial deviations from $\varepsilon=-1/3$ are expected to occur only for systems with a black hole mass exceeding a few times $10^8 M_\odot$ (see equation \ref{eq:M_star}).

As the secondary loses mass, the GW and Kelvin-Helmholtz timescales change. If the secondary's remaining mass is \(M_2' \leq M_2\), the timescales ratio changes as
\begin{equation}
    \frac{\tau_{KH}}{\tau_{GW}} \approx  \left. \frac{\tau_{KH}}{\tau_{GW}} \right|_{MS} \left(\frac{M_2'}{M_2}\right)^{4/3-5\varepsilon_{ad}}\,,
\end{equation}
with \(\left. \frac{\tau_{KH}}{\tau_{GW}} \right|_{MS}\) given by equation \ref{eq:timescales_ratio_MS}, being greater than unity in the discussed regime.
Since \(\varepsilon_{ad}<0\), the power \((4/3-5\varepsilon_{ad})\) is positive, and thus the ratio \(\tau_{KH}/\tau_{GW}\) approaches unity as \(M_2'\) decreases in the process of secondary mass loss. After losing sufficient mass, the two timescales therefore become similar.

\subsubsection{Equal timescales evolution - $\tau_{GW}=\tau_{KH}$}

As explained above, in either initial regime, the GW and Kelvin-Helmholtz timescales become similar after the secondary has lost sufficient mass. Therefore, once the timescales become similar, the system will continue to evolve such that the condition $\tau_{GW}=\tau_{KH}$ is maintained. Given the dependence of \(\tau_{KH}\) and of \(\tau_{GW}\) on the secondary's mass and radius, we find that the mass-radius exponent which yields equal timescales is roughly \(\varepsilon_{eq}=4/15\), where we assumed constant opacity in equation \ref{eq:tau_KH_kappa}.

The evolution timescale in this regime, as a function of the remaining secondary's mass is given by
\begin{equation} \label{eq:tau_eq_constant_opacity}
    \tau_{GW} = \tau_{KH} \approx 1.3\times10^6 \left(\frac{M_2}{M_{\star}}\right)^{-19/15} \left(\frac{M_1}{M_{SgrA*}}\right)^{-2/3} \rm yr \,.
\end{equation}

Since the secondary's luminosity is also given by $L\approx 4\pi R_2^2 \sigma T^4_{eff}$, the cooling timescale can be written as
\begin{equation}
    \tau_{KH} \propto \frac{M^2}{T_{eff}^4 R_2^3} \,,
\end{equation}
and hence the surface temperature scales with the secondary's mass as $T_{eff} \propto M_2^{37/60}$, when $\tau_{GW}=\tau_{KH}$. We conclude that as the secondary loses mass and evolves along this track, its surface temperature gradually decreases. When the surface is sufficiently cold ($T_{eff} \lesssim 4000 \rm K$), the $H^-$ absorption at the surface becomes the dominant source of opacity. Due to the high surface opacity, radiative transfer becomes inefficient, and the secondary becomes entirely convective. The strong dependence of the $H^-$ opacity on temperature implies that the secondary maintains approximately constant $T_{eff}$, with the luminosity depending mostly on the secondary's radius. We therefore obtain the late time evolutionary track of the mass losing secondary, which is also characterized by \(\tau_{GW}=\tau_{KH}\), but with a constant $T_{eff} \approx 4000~{\rm K}$. This track is reminiscent of the Hayashi line appearing in the Hertzsprung-Russell diagram \citep[e.g.][]{Kippenhahn_1994}.

Evolving with $\varepsilon_{eq}$, the surface temperature decreases to the Hayashi line value, $4000 \rm K$ when the secondary's mass is $M_2=M_{hl}$, where
\begin{equation} \label{eq:M_hl}
    M_{hl} = 1.1 M_\odot \left( \frac{M_1}{M_{Sgr A*}} \right)^{0.1} \,.
\end{equation}

Assuming that the surface temperature remains constant at lower masses, the radius evolves with mass with $\varepsilon_{hl}=13/21$ for masses smaller than $M_{hl}$. The longest evolution time is achieved near the transition to the constant surface temperature regime, at $M_2=M_{hl}$, and is roughly
\begin{equation}
    \tau_{max} \approx 10^7 \left( \frac{M_1}{M_{Sgr A*}} \right)^{-0.47} \rm yr\,.
\end{equation}

\begin{figure}
\centering
\includegraphics[trim={1cm 1cm 1cm 0 }, width=\columnwidth]{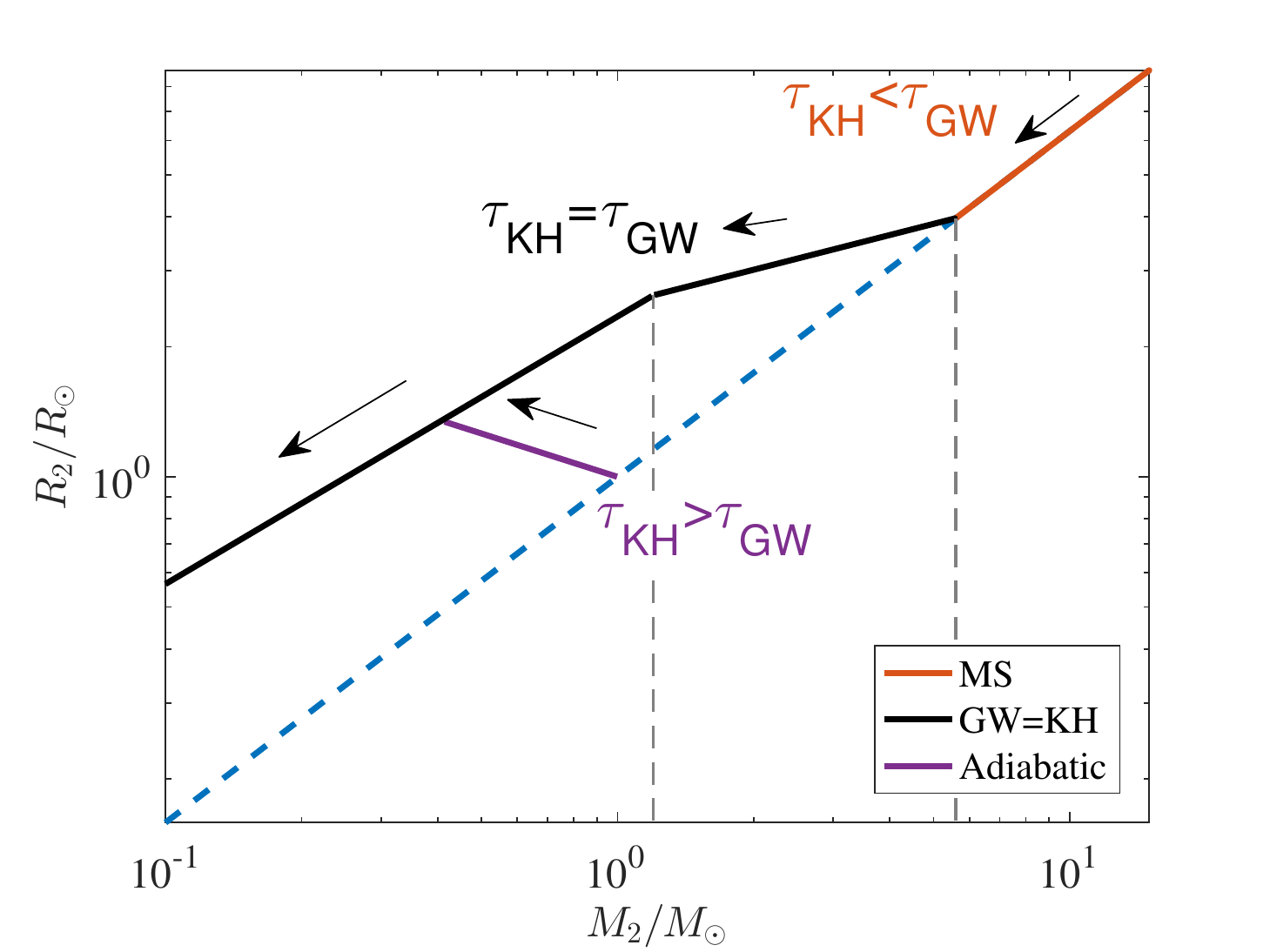}
\caption{The secondary's radius as a function of its mass, during stable mass transfer. The primary is taken to be \(M_1 = 4\times10^6 M_{\odot}\). \textit{Dashed blue line} represents the mass-radius relation of main-sequence stars (equation \ref{eq:MassRadiusMS}). \textit{Red line} shows the evolution of massive stars with $M_2>M_{\star}$, where GW timescale is longer than the Kelvin-Helmholtz timescale. The star then follows the main sequence as it loses mass, until reaching a mass of $M_\star$ (equation \ref{eq:M_star}). If the star's initial mass is sufficiently low, $M<M_\star$, initially $\tau_{KH}>\tau_{GW}$ as mass transfer ensues. The mass-radius relation deviates from that of the main-sequence, and the secondary evolves with $\varepsilon_{ad}$. The \textit{purple line} exemplifies such evolution for a secondary's initial mass of $1 M_{\odot}$. The initial evolution, in either regime, eventually leads to $\tau_{GW}=\tau_{KH}$. Later evolution follows the \textit{black curve} corresponding to $\varepsilon_{eq}=4/15$ for remaining mass higher than $M_{hl}$ (equation \ref{eq:M_hl}), and eventually when $M_2<M_{hl}$ the secondary maintains a constant surface temperature, with $\varepsilon_{hl}=13/21$. The \textit{dashed grey lines} mark the transition masses, $M_\star \approx 5.6 M_\odot$ and $M_{hl}\approx 1.1 M_\odot$.}
\label{fig:MassRadius_evolution}
\end{figure}

We summarize the discussion on the different types of evolution under mass transfer and their corresponding value of \(\varepsilon\). If the initial mass of the star, when it has first reached its Roche limit is sufficiently high, $M_2>M_{\star}$ then \(\varepsilon=\varepsilon_{MS}\). The star then loses mass while evolving as a main sequence star until reaching a mass of $M_2=M_{\star}$ where \(\tau_{GW}\) and \(\tau_{KH}\) become similar. From that point the star evolves maintaining \(\tau_{GW}=\tau_{KH}\). When the surface temperature is sufficiently high, for $M_2>M_{hl}$, \(\varepsilon=\varepsilon_{eq}\). As the mass decreases below $M_{hl}$, the secondary becomes fully convective due to the high surface opacity, and evolves with $\varepsilon=\varepsilon_{hl}$, maintaining both $\tau_{GW}=\tau_{KH}$ and $T_{eff}\approx const$. Alternatively, if the initial mass is lower $M_2<M_{\star}$, the star initially loses mass while evolving adiabatically, with \(\varepsilon=\varepsilon_{ad}\). Once the thermal and GW timescales become similar, it evolves with either $\varepsilon_{eq}$ or $\varepsilon_{hl}$, similarly to the track followed by high mass stars. These results are presented in figure \ref{fig:MassRadius_evolution} and in the following equation:
\begin{equation}
    \varepsilon = \left\{\begin{array}{ll}
        \varepsilon_{MS} \approx 0.8, & \text{initially for } M_2>M_\star \\
        \varepsilon_{ad} = \frac{2-\gamma}{4-3\gamma}, & \text{initially for }  M_2<M_\star\\
        \varepsilon_{eq} = 4/15, & \text{late time evolution, } M_{hl} < M_2 < M_\star \\
        \varepsilon_{hl} = 13/21, & \text{late time evolution, } M_2 < M_{hl}
        \end{array} \right.
\end{equation}

The results we obtained in this section for the mass-radius evolution are somewhat crude. For example, we neglected the possible transition to Kramer's opacity which may dominate in lower temperatures and higher densities, or the contribution of degeneracy pressure. In addition, we treat the star's interior as fully radiative in order to derive its luminosity, ignoring its more intricate structure. However, our model does capture some of the key ingredients of the problem at hand. The first important observation is that the radius of the star is expected to evolve differently depending on whether its mass loss rate is faster or slower than its cooling time. This distinguishes between the adiabatic evolution (when $\tau_{GW} \ll \tau_{KH}$), and the main-sequence evolution (where $\tau_{KH} \ll \tau_{GW}$). Additionally, in later times, the system eventually evolves while maintaining $\tau_{GW} \approx \tau_{KH}$. The condition for sustaining $\tau_{GW} \approx \tau_{KH}$ sets the mass-radius relation, assuming that the luminosity's dependence on mass and radius, $L(M_2,R_2)$ is given. Here we considered either constant opacity and radiative structure (yielding $\varepsilon_{eq}$), and for lower masses, high surface opacity with convective structure, resulting in roughly constant surface temperature (from which we calculated $\varepsilon_{hl}$).

\subsection{Extracting $\varepsilon$ from the GW signal} \label{subsec:epsilon_from_GW}

In this subsection we demonstrate how \(\varepsilon\) can be inferred from the GW signal emitted by the system. Similar calculations in the context of interacting double white dwarf binaries appear in \cite{Nelemans_2004_AMCVn_GW}, and earlier in \cite{Faulkner_71,Vila_71}, discussing evolution of cataclysmic binary systems. The GW frequency changes with orbital separation as \(f \propto a^{-3/2}\) (equation \ref{eq:f_GW_gen})
\begin{equation}
    \frac{\dot{f}}{f} = -\frac{3}{2} \frac{\dot{a}}{a} \,.
\end{equation}

Another observable property of the gravitational wave signal is the strain amplitude, which, as in equation \ref{eq:h_GW_gen}, depends inversely on the orbital separation, and linearly on the secondary's mass, \(h \propto M_2 a^{-1}\), so that
\begin{equation}
    \frac{\dot{h}}{h} = \frac{\dot{M_2}}{M_2} - \frac{\dot{a}}{a} = \left(\frac{\frac{4}{3} - \varepsilon}{\varepsilon - \frac{1}{3}}\right) \frac{\dot{a}}{a} \,,
\end{equation}
where we used equation \ref{eq:stab_mt_adot_mdot} in the second equality.

The value of \(\varepsilon\) can be therefore extracted from the following ratio:
\begin{equation} \label{eq:K_GW}
    K_{GW} \equiv \frac{\dot{f}/f}{\dot{h}/h} = \frac{\frac{1}{3} - \varepsilon}{\frac{8}{9} - \frac{2}{3}\varepsilon} \,,
\end{equation}
or alternatively, rearranging:
\begin{equation}
    \varepsilon = \frac{3-8K_{GW}}{9-6K_{GW}} \,.
\end{equation}

The measurement of \(K_{GW}\) is robust due to its lack of sensitivity to any detector-specific calibrations, and in particular to the precise value of \(h\), which is prone to uncertainties such as the detector's angular response. "Standard" GW evolution, without mass transfer, as in binary systems outside the Roche limit, obeys \(\dot{h}/h = - \dot{a}/a\) and consequently \(K_{GW} = 3/2\). The presence of mass transfer modifies this dimensionless number. 

Massive stars, $M_2>M_\star$ which initially evolve with \(\tau_{KH} \ll \tau_{GW}\), have \(\varepsilon=\varepsilon_{MS}\) and  would therefore display GW signals with $K_{GW}=-1.3$.
Less massive stars, initially evolve with \(\tau_{GW}\ll \tau_{KH}\) and evolve adiabatically. For \(\gamma = 5/3\), $\varepsilon=-1/3$ and therefore \(K_{GW}=3/5\). Later in the evolution, and for all initial masses, the system evolves with \(\tau_{GW}=\tau_{KH}\). When $\varepsilon_{eq}=4/15$, we have \(K_{GW}=3/32\), and ultimately when $\varepsilon_{hl}=13/21$, $K_{GW}=-3/5$.

$K_{GW}$ is a robust dimensionless number that easily distinguishes between interacting and non interacting binaries which evolve through GW radiation. If mass transfer occurs, it reveals which of the possible evolutionary tracks is taken.

\subsection{The Reverse Chirp}

One of the unique features of the gravitational wave emission associated with these mass transferring systems, when \(\varepsilon<1/3\), is the \textit{decrease} in strain amplitude and frequency with time. Prior to the onset of mass transfer, the emission of gravitational waves from a binary system tends to decrease the orbital separation, thus increasing the orbital frequency and the strain amplitude. In the presence of stable mass transfer from a component with sufficiently small \(\varepsilon\), the orbital separation of the binary increases, and the GW signal will be characterized by a \textit{``reverse-chirp"}. Therefore, a GW signal whose frequency and strain amplitude decrease in time is likely to originate from a binary system in which stable mass transfer takes place.

For mass transferring systems with \(\varepsilon>1/3\) the frequency and strain amplitude increase with time, as in a regular chirp, but still produce a unique value of \(K_{GW}\), which can be used for detection of systems with mass transfer taking place (equation \ref{eq:K_GW}). 

For high initial secondary mass $M_2>M_\star$, the secondary's radius changes with \(\varepsilon_{MS}\approx 0.8 > 1/3\). At this stage the system evolves with a regular chirp, as the binary separation decreases with mass loss. After losing enough mass, $M_2 \leq M_\star$, and we have \(\varepsilon_{eq}=4/15 < 1/3\). Therefore, at this regime the binary separation increases with time, and we have a reverse chirp. This critical mass marks the transition between the regular and reverse chirp of the GW signal emitted by the binary system. As $M_2$ decreases below $M_{hl}$, we have $\varepsilon_{hl}=13/21>1/3$, and the system evolves with a regular chirp again. If alternatively the initial mass is small, $M_2<M_\star$, the star's radius initially evolves with \(\varepsilon_{ad} < 1/3\), producing a reverse chirp. After losing sufficient mass the system transitions to \(\varepsilon_{hl}\), perhaps after some transient phase for which $\varepsilon=\varepsilon_{eq}$, characterized by a reverse chirp. Thus, late enough in the evolutions of all systems, the GW signal will be of a forward chirp.

\cite{Dai_Blandford_2013} have also discussed stable mass transfer between a star and a SMBH, and argued that the star evolves adiabatically and the orbit expands. 
Our results are consistent with theirs only for low mass stars, $M_2<M_\star$. For more massive stars we show that the cooling time is short compared to the evolutionary time under gravitational waves so the stars evolve on the main sequence, and the orbit shrinks.
Our conclusions regarding the final state of the binary also differ. \cite{Dai_Blandford_2013} noted that eventually the mass loss rate decreases, until the star begins to cool down. As described in our work, this indeed occurs when \(\tau_{GW} \approx \tau_{KH}\). However, Dai \& Blandford argue that at this stage mass transfer ceases, and the star "parks" at that orbit. They also suggest that in later times the orbit shrinks again due to GW emission, until mass transfer is resumed, and the orbit might expand again.
In contrast, we show that mass transfer does not cease when the cooling time becomes comparable to the GW timescale. Rather than parking, the orbit continues to evolve, and the secondary's radius is set by maintaining \(\tau_{GW} \approx \tau_{KH}\). Finally, our works also differ in the criterion for the onset of this \(\tau_{GW} \approx \tau_{KH}\) stage. We have calculated explicitly how the cooling rate of the star evolves as it losses mass, rather than taking the initial Kelvin-Helmholtz time of the star, before the onset of mass transfer, as in their paper.

Previous theoretical works on double white dwarf binaries have discussed their evolution due to the emission of gravitational waves \citep{MNS_DWD_MT,SVN_DWD_GW}. It was found that in double white dwarf binaries with sufficiently small mass ratio, stable mass transfer between the components will occur eventually, as the orbit shrinks. In such binaries, the orbital period will increase with time as mass transfer takes place, similarly to the reverse-chirp scenario discussed here. The GW frequency emitted by these systems will be within the sensitivity range of space interferometer GW detectors like eLISA \citep{eLISA_2012,eLISA_2013}. A few known interacting double white dwarf binaries will serve as verification sources for eLISA's performance, as they are expected to be detected shortly after the beginning of the mission \citep{Nelemans_2004_AMCVn_GW}. Interacting white dwarf binaries differ from the systems studied in this work, by evolving with a single value of \(\varepsilon\), usually taken to be \(\varepsilon=-1/3\). In the systems we consider, which contain main sequence stars, the secondary follows different values of \(\varepsilon\) as the system evolves, corresponding to different evolutionary regimes. In addition, binary white dwarfs usually have mass ratios of order unity, which can affect the stability of mass transfer \citep{MNS_DWD_MT}. On the contrary, in our work, $q \ll 1$, and mass transfer stability is independent of $q$. 

The \textit{braking index}, defined as \(n_{br} \equiv \ddot{f}f/\dot{f}^2\) is an additional quantity which can be used to discriminate the driving mechanism of the orbital evolution. Since the frequency evolution occurs on a timescale \(\tau_{GW}\), we have
\begin{equation} \label{eq:f_dot_gen}
    \dot{f} \approx f/\tau_{GW} \propto f a^{-4} M_2 \,,
\end{equation}
where we used equation \ref{eq:GW_timescale_gen}. Note that the in the limit of \(q \ll 1\), \(M_1\) is taken to be constant.

Rewriting equation \ref{eq:f_dot_gen}, and expressing \(a\) and \(M_2\) through \(f\) using equations \ref{eq:RL_size}, \ref{eq:MassRadius} and \ref{eq:f_GW_gen} we have
\begin{equation}
    \dot{f} \approx f/\tau_{GW} \propto f^{\frac{11}{3} +  \frac{2}{1-3\varepsilon}} \,.
\end{equation}

While for a non-interacting system driven by the emission of gravitational radiation, in the absence of mass transfer \(n_{br}=11/3\), here the braking index will be:
\begin{equation} \label{eq:BreakingIndex_MT}
    n_{br} = \frac{11}{3} + \frac{2}{1-3\varepsilon} \,.
\end{equation}

As with the measurement of \(K_{GW}\), the value of \(n_{br}\) can be used for inferring \(\varepsilon\). In mass transferring systems, which eventually evolve with similar cooling and GW timescales, the breaking index will be \(n_{br}=4/3\) sufficiently late in their evolution, corresponding to $\varepsilon_{hl}=13/21$.

\subsection{L2 mass loss and its consequences} \label{subsec:L2_MT}
We now examine the conditions for mass loss through L2 in systems of a main-sequence star orbiting a SMBH. Main-sequence stars with radiative exteriors (\(M_2 \gtrsim M_{\odot}\)) have an outer density profile that can be modelled using an \(n=3\) polytrope. Low mass (\(M_2 \lesssim M_{\odot}\)) main-sequence stars have convective envelopes, which follow an \(n=1.5\) polytropic profile. As we found in section \ref{sec:GW_Evolution}, L2 mass loss is made possible when \(n>1.5\). We therefore expect L2 mass loss to occur only in stars with a mass higher than \(M_{\odot}\).

According to the criterion in equation \ref{eq:GW_LeakageCriterion}, L2 mass loss occurs within a certain mass range. Using the dependence of \(\beta\) on \(M_2\) (equation \ref{eq:MS_beta}), we have the following limits on the star's mass:
\begin{equation} \label{eq:ISCO_tidal}
    \frac{M_2}{M_{\odot}} < 24 \left(\frac{M_1}{M_{Sgr A*}}\right)^{1.3} \,,
\end{equation}

\begin{equation} \label{eq:L2_leakage_cond}
    \frac{M_2}{M_{\odot}} > \max \left\{ 1 , (2\times10^{-3}) \left(\frac{M_1}{M_{Sgr A*}}\right)^{1.43} \right\} \,.
\end{equation}

Therefore, for a SMBH primary with the mass of Sgr A*, L2 leakage occurs for stars with a mass between \(1\) to \(24\) solar masses. If this L2 mass ejection is substantial compared to the L1 mass loss, the evolution timescale of the GW signal will be altered significantly, as we now show.

Assuming that the black hole's mass and distance are known, the orbital separation and secondary's mass can be deduced from the GW frequency and strain amplitude, \(f\) and \(h\) (equations \ref{eq:f_GW_gen} and \ref{eq:h_GW_gen}). The timescale of angular momentum loss due to GW emission, \(\tau_{GW} \equiv -(J/\dot{J}_{GW})\), is then given by equation \ref{eq:GW_timescale_gen}.

In the absence of mass transfer, the GW frequency changes on a timescale given by
\begin{equation}
    \tau_f^{NoMT} \equiv \frac{f}{\dot{f}} = \tau_{GW}/3
\end{equation}

If mass transfer occurs, the frequency evolution timescale is given by (using equations \ref{eq:stab_mt_adot_mdot} and \ref{eq:Mdot_Jdot})
\begin{equation} \label{eq:f_timescale_MT}
    \tau_f = \left( \frac{5/3+\varepsilon-2\alpha}{\varepsilon-1/3} \right) \tau_f^{NoMT}\,,
\end{equation}
where \(\alpha\) is the fraction of mass loss that is ejected from L2 (as defined in section \ref{sec:MT_stab}). We emphasize that \(K_{GW}\) is independent of \(\alpha\), and so \(\varepsilon\) can be inferred from the GW signal with or without L2 mass loss.

The term in parentheses in equation \ref{eq:f_timescale_MT} is the deviation from the detached evolution timescale. Note that it may be negative (for \(\varepsilon<1/3\)) corresponding to a reverse-chirp evolution. This term is also in agreement with the factor appearing in equation 12 in \cite{Nelemans_2004_AMCVn_GW}, with \(\alpha = 0\), as they did not consider a non-conservative component of the mass transfer.

The effect of the (non-conservative) mass transfer through L2 as compared to conservative mass transfer through L1 alone is thus given by
\begin{equation}
    \frac{\Delta \tau_f}{\tau_f} = \frac{\alpha}{5/6+\varepsilon/2} \,.
\end{equation}

As previously discussed, the value of \(\varepsilon\) depends on the comparison between the GW timescale to the Kelvin-Helmholtz timescale of the star. For a fixed black hole, there is a threshold secondary mass, $M_\star$ for which \(\tau_{GW} \approx \tau_{KH}\), marking the transition between the different values of \(\varepsilon\) (equation \ref{eq:M_star}).

If mass transfer through L2 occurs, the value of \(\alpha\) can be estimated using equations \ref{eq:ML2_ML1_gen} and \ref{eq:depth_ratio_gw}, giving
\begin{equation}
    \frac{\alpha}{1-\alpha} =\dot M_{L_2}/\dot M_{L_1} = \left(1-q^{\frac{13}{27}} \beta^{-\frac{10}{9}}\right)^{9/2} \,,
\end{equation}
where we used \(n=3\). If the expression in parentheses is negative, there is no mass transfer through L2 and $\alpha=0$.

We express \(\alpha\) as a function of \(M_2\) by using equation \ref{eq:MS_beta} and get
\begin{equation}
    \frac{\alpha}{1-\alpha} = \left(1- \beta_{\odot}^{-\frac{10}{9}} \left(\frac{M_1}{M_{\odot}}\right)^{-\frac{13}{27}}  \left(\frac{M_2}{M_{\odot}}\right)^{0.37}\right)^{9/2} \,.
\end{equation}

\begin{figure}
\centering
\includegraphics[trim={1cm 0 1cm 0 }, width=\columnwidth]{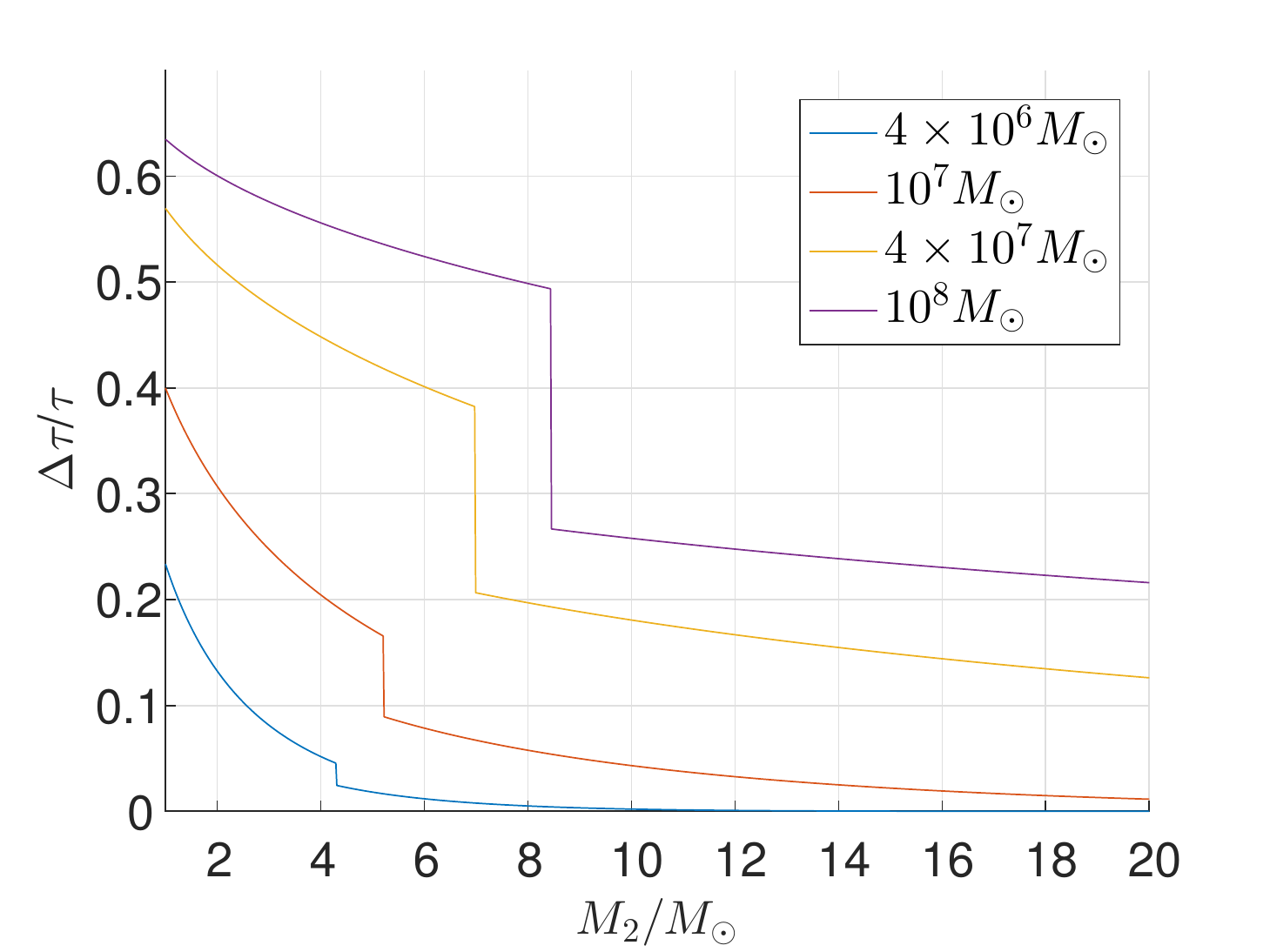}
\caption{Stellar EMRI - the relative change in the GW signal evolution timescale due to mass loss through the L2 Lagrange point. The system consists of a SMBH of mass and a main-sequence star of mass \(M_2\). Different values of \(M_2\) correspond to different rates of mass loss through L2, leading to varying evolution timescales of the frequency and strain amplitude of the GW signal. Different curves show the change in evolution timescales as a function of secondary mass, for different SMBH mass measured in solar mass, as shown in the legend. The discontinuity in the curve originates in the transition in the mass-radius exponent of the star when the GW timescale becomes similar to the Kelvin-Helmholtz time of the star, occurring at $M_2=M_\star$.}
\label{fig:DeltaT_M2}
\end{figure}

The value of the relative timescale correction is plotted against the secondary's mass in figure \ref{fig:DeltaT_M2}. The discontinuity arises from the transition between two possible values of \(\varepsilon\), \(\varepsilon_{MS}\) or \(\varepsilon_{ad}\). We find that for common main-sequence star masses around \(M_2 \approx 2 M_{\odot}\), the signal frequency and strain amplitude will proceed on a timescale which is shorter by a few tens of percent, compared to the expected timescale when L2 leakage is ignored.

Failure to consider the effects of mass transfer on the orbital evolution can also lead to incorrect interpretation of the system's \textit{chirp mass}. The chirp mass of a binary system is given by \citep[e.g.][]{Blanchet_1996}
\begin{equation} \label{eq:ChirpMass_noMT}
    \mathcal{M}_{ch} = \frac{(M_1M_2)^{3/5}}{(M_1+M_2)^{1/5}} \,.
\end{equation}

When the binary evolves just through the emission of GW, the chirp mass can be calculated from the GW frequency and frequency time derivative, \(f\) and \(\dot{f}\), and is given by \citep[e.g.][]{LIGO_detection}
\begin{equation} \label{eq:ChirpMass}
\mathcal{M}_{ch} = \frac{c^3}{G} \left[\frac{5}{96} \pi^{-8/3} f^{-11/3} \dot{f}\right]^{3/5} \,.
\end{equation}

This relation is being used for constraining the masses of the binary components, when only the GW signal is available, as done in the recent LIGO detections. However, when mass transfer takes place, equation \ref{eq:ChirpMass} should be replaced by
\begin{equation} \label{eq:ModifiedChirpMass}
\mathcal{M}_{ch} = \frac{c^3}{G} \left[ \left( \frac{\varepsilon-1/3}{5/3+\varepsilon-2\alpha} \right) \frac{5}{96} \pi^{-8/3} f^{-11/3} \dot{f}\right]^{3/5} \,.
\end{equation}

Incorrectly using equation \ref{eq:ChirpMass} when mass transfer occurs will yield chirp mass which is wrong by a factor of order unity. Specifically, in the case of reverse chirp scenario (i.e, \(\varepsilon<1/3\)), equation \ref{eq:ChirpMass} gives a complex result. In the case of the forward chirp (\(\varepsilon=\varepsilon_{MS}=0.8\)), equation \ref{eq:ChirpMass} deviates from the physical chirp mass by a factor of 0.37-0.5, when \(\alpha\) varies between between \(0\) and \(1/2\).

\subsection{Initial system evolution}

\begin{figure}
\includegraphics[trim={1cm 1cm 1cm 1cm },width=\columnwidth]{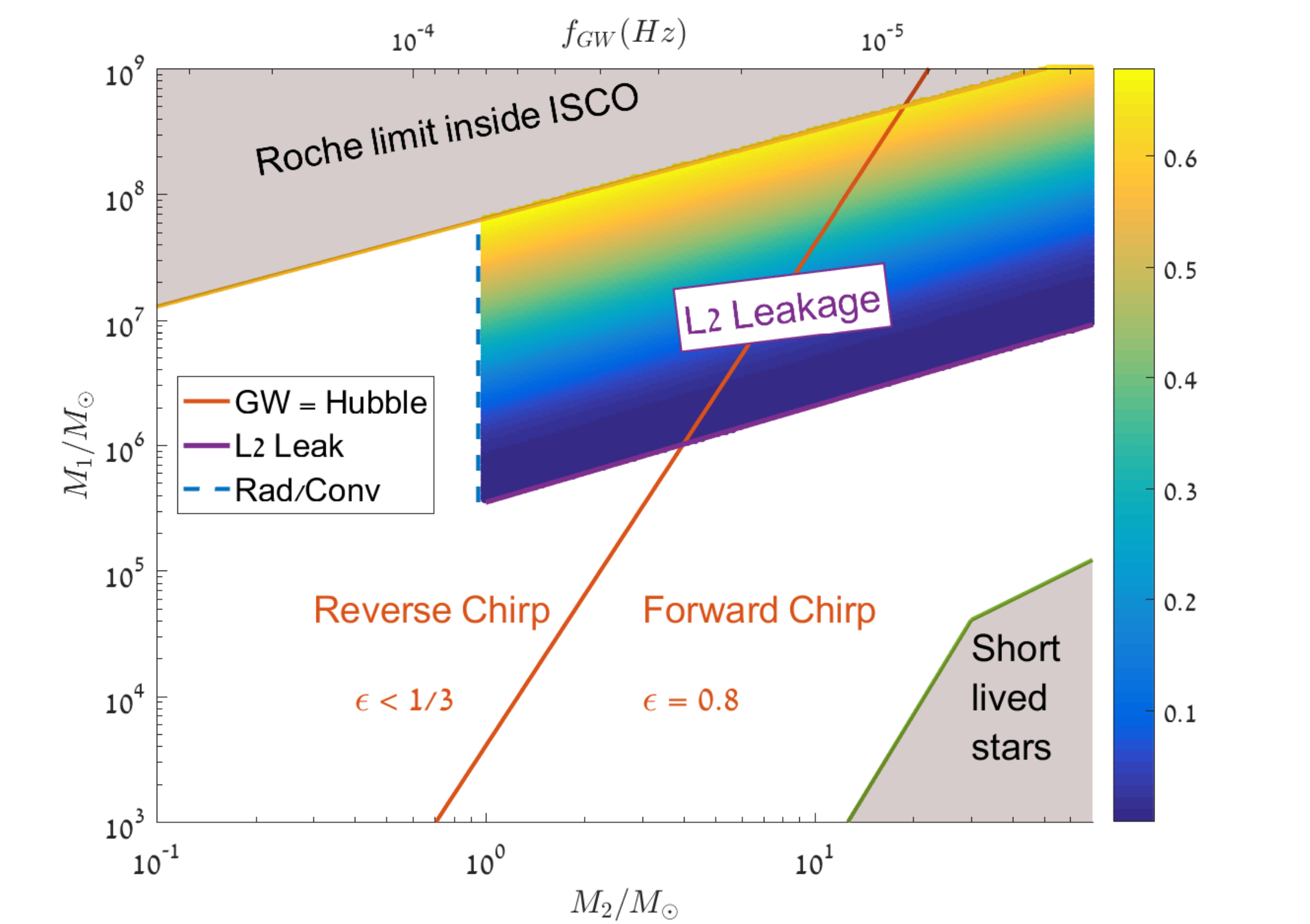}
\caption{The mass transfer landscape. Vertical axis is the BH mass. Horizontal axis is the star's mass (bottom) or GW frequency (top). \textit{Shaded areas} represent regions of non-interest. Along the \textit{orange} line \(\tau_{GW}\) equals the Kelvin-Helmholtz timescale of a main-sequence star of mass \(M_2\). The \textit{dark yellow} line shows where the Roche limit occurs on the horizon of the black hole. The region above the \textit{purple} line is where L2 leakage occurs. The star's lifetime equals \(\tau_{GW}\) along the \textit{green} line. It breaks at around $M_2=30 M_\odot$ since the lifetime of very massive stars is a few million years, independently of the stellar mass, whereas the lifetime of lighter stars changes with mass. The envelope of main sequence stars transitions from convective to radiative on the \textit{dashed} line. The colour in the L2 leakage regime indicates the ratio \(|\dot{M}_{L_2}/\dot{M}_{L_1}|\), as given by the colour bar on the right-hand side.}
\label{fig:MassMap}
\end{figure}

We consider the various binary system's evolutionary regimes at the onset of mass transfer assuming a secondary of mass $M_2$ which starts transferring mass as a main-sequence star, orbiting a massive black hole of mass $M_1$. At this instance, the system is represented by a point (\(M_2,M_1\)) on figure \ref{fig:MassMap}. The secondary is a main-sequence star, and its radius is known from main-sequence relations (equation \ref{eq:MassRadiusMS}). Since the star fills it Roche lobe, its distance from the SMBH is known (equation \ref{eq:a_MS_SMBH}) and therefore the GW frequency and its evolution timescale, \(\tau_{GW}\), are given by equations \ref{eq:tau_GW_MS_SMBH} and \ref{eq:f_GW_EMRI}. The GW frequency, for each secondary mass \(M_2\), is independent of the SMBH mass, and is shown on the top horizontal axis of figure \ref{fig:MassMap}. As the secondary loses mass, and the black hole's mass remains roughly constant, the system moves to the left on figure \ref{fig:MassMap}.

If the GW evolution is initially sufficiently slow compared to the cooling rate of the star through radiation (i.e. \(\tau_{KH} \ll \tau_{GW}\)), the star  continues to follow the mass-radius relation of main-sequence stars as it loses mass (equation \ref{eq:MassRadiusMS}). In this case, the secondary evolves and loses mass while remaining a main-sequence star, and simply moves to the left on the figure. The GW signal emitted by such systems is characterized by a "forward chirp", since \(\varepsilon_{MS} \approx 0.8 > 1/3\). This evolutionary phase ends once the cooling time becomes comparable to the GW timescale, when $M_2 = M_\star$ as given by equation \ref{eq:M_star}, and represented by the \textit{orange} line on figure \ref{fig:MassMap}.

As the star continues to lose mass beyond $M_\star$, its radius adjusts so \(\tau_{GW} = \tau_{KH}\) is maintained. This is achieved by a mass-radius exponent of \(\varepsilon_{eq}=4/15\) in the constant opacity regime. Note that since \(\varepsilon_{eq}<1/3\) the system produces a ``reverse chirp", with its frequency and strain amplitude decreasing in time. 
As the remaining secondary mass decreases, $M_2<M_{hl}$, its mass-radius exponent is $\varepsilon_{hl}=13/21$, corresponding to constant surface temperature. This results in a ''forward chirp", as the orbit shrinks with mass loss.
This evolutionary regime, for which $\tau_{GW}=\tau_{KH}$ is described in figure \ref{fig:MassRadius_evolution}. Since the secondary at this point deviates from the main-sequence, the system does not correspond to a point on \ref{fig:MassMap}.

On the other hand, consider a system for which initially \(\tau_{GW} \ll \tau_{KH}\). This occurs when the secondary's initial mass before the onset of mass transfer, is sufficiently small, $M_2<M_\star$. Here the system evolves with \(\varepsilon_{ad}\), which is generally negative. The star inflates as it moves further out from the black hole, resulting in a reverse chirp.
In this regime, the ratio \(\tau_{KH}/\tau_{GW}\) decreases towards unity as the star loses mass. When the two timescales become similar, the value of \(\varepsilon\) changes, so that \(\tau_{GW} = \tau_{KH}\) is maintained. This can be seen in figure \ref{fig:MassRadius_evolution}, for a fixed black hole mass. Here again, the secondary deviates from the main-sequence, and figure \ref{fig:MassMap} only describes the initial evolution.

We exclude some regions in figure \ref{fig:MassMap} which are irrelevant for our analysis. First, the GW timescale must be shorter than the lifetime of the star. The region below the \textit{green} curve in figure \ref{fig:MassMap} is therefore excluded. We estimate the main-sequence lifetime as \(\tau_{nuc} \propto M_2 c^2 / L\). For extremely massive main-sequence star, $L$ approaches the Eddington luminosity, and thus $\tau_{nuc}$ is constant (a few million years). In less massive stars, the lifetime varies with mass roughly as $\tau_{nuc} \propto M_2^{-2}$. Note that for the entire range of massive black holes we consider, $M_1>10^3 M_\odot$, the GW timescale is shorter than the Hubble time.


In section \ref{sec:GW_Evolution}, we derived the condition for the Roche limit of the secondary to be outside the horizon of the black hole. This condition defines the region in the graph for which mass transfer might occur - outside this region, stars never overfill their Roche lobe before the Schwarzschild radius, and plunge directly into the event horizon. This sets the shaded region above the \textit{dark yellow} line in figure \ref{fig:MassMap}.

We also derived the conditions for leakage through L2 to occur. We found that the density profile power law at the surface of the star, \(n\), must be greater than \(1.5\). For low mass stars (\(<M_{\odot}\)), the envelope is convective, and follows a power law of \(n=1.5\). However, for stars of mass \(M_2 > M_{\odot}\), the envelope is radiative, and \(n=3\). These conditions set an area in the graph in which some mass is lost through L2. Within this parameter space region, the amount of mass lost through L2 compared to the mass transferred through L1 varies. The \textit{dark yellow} line, above which no mass transfer occurs, and the \textit{purple} line which marks the onset of L2 mass loss have similar slopes, as can be seen in equations \ref{eq:ISCO_tidal} and \ref{eq:L2_leakage_cond}.

\section{Summary and Discussion} \label{sec:Discussion}

Roche lobe overflow occurring during stable mass transfer is an essential ingredient in numerous astrophysical phenomena. Usually, stable mass transfer is assumed to occur only through the inner Lagrange point, L1. In this work, we have extended the classical formalism of Roche lobe overflow, to calculate mass leakage occurring through the outer Lagrange point, L2. We have shown that such leakage occurs for binary systems with sufficiently fast orbital evolution, and an extreme mass ratio.

In section \ref{sec:MT_stab} we found that for extreme mass ratios, mass transfer through L2 makes the mass transfer less stable, but is not usually sufficient to make the system dynamically unstable. For a wide range of the secondary mass-radius relations ($\varepsilon>-2/3$), mass transfer through L2 does not lead to loss of stability. The mass transfer rate through L2 is always smaller than the corresponding L1 mass flow because the Roche potential at L1 is always lower than that at L2. Our stability analysis neglects other possible channels of orbital angular momentum loss through the mass transfer, such as spin up of the primary, or additional torques of the outflow applied on the orbiting secondary.

The L2 mass transfer was calculated for arbitrary external driving mechanism. We have found the specific conditions for L2 mass loss to occur in binary systems driven by the emission of gravitational radiation. We then proceeded to study the mass transfer occurring at a system of a main-sequence star orbiting a SMBH at its Roche limit. The Roche limit in these systems can be merely a few times the horizon of the SMBH, thus resulting in strong GW emission. A Roche-lobe filling star orbiting the SMBH in our Galactic Centre, produces GW signals detectable by future GW detectors such as eLISA. A similar system of a star orbiting the SMBH at the centre of the Andromeda galaxy may be also detectable, but with a slightly lower signal to noise ratio. The period of these gravitational waves is similar to the dynamical time of the secondary, of order hours.

The GW emission associated with such binaries evolves with time, as the masses and orbit change. The GW temporal evolution differs from the characteristic chirp signal of two coalescing point masses, due to the presence of stable mass transfer between the components. This stable mass transfer results in a coordinated evolution of the orbital separation and the size of the secondary. The radius evolution of the mass losing star, along with the amount of mass lost through L2, affect fundamental properties of the GW signal. For a GW signal with frequency $f$ and strain amplitude $h$, we define $K_{GW}\equiv (f\dot{h})^{-1}\dot{f} h$. $K_{GW}$ is a robust dimensionless number, which can be used to determine whether the source of the gravitational radiation is a mass transferring binary system, or a non-interacting binary system. For non interacting binaries, we have $K_{GW}=3/2$, while if mass transfer occurs, the value of $K_{GW}$ reveals the mass-radius relation of the secondary (see equation (\ref{eq:K_GW})). This can also be achieved using the signal's braking index, $n_{br} \equiv \ddot{f}f/\dot{f}^2$, as in equation \ref{eq:BreakingIndex_MT}.

In \S\ref{subsec:MassRadius_EMRI} we studied evolutionary tracks of the secondary's radius as it loses mass. This evolution mainly depends on the comparison between the mass loss rate, given by the GW timescale, $\tau_{GW}$, and the secondary's cooling time, $\tau_{KH}$. For a sufficiently massive main-sequence star, $\tau_{KH} \ll \tau_{GW}$ and the star's size adjusts according to the standard main-sequence mass-radius relation. These two timescales become similar at some critical mass ($M_\star$, see equation \ref{eq:M_star}), where the secondary's radius begins to deviate from the main-sequence radius.

Main-sequence stars with low initial mass ($M_2<M_\star$) have cooling times longer than their mass-loss rate, $\tau_{GW} \ll \tau_{KH}$. The secondary then adjusts adiabatically, increasing its  radius as its mass decreases. After losing enough mass, the two timescales become similar. Later evolution (for any initial mass) follows $\tau_{GW} \approx \tau_{KH}$. Depending on the assumed model for the luminosity of the secondary at this point, the mass-radius evolution is derived. We found two evolutionary regimes maintaining $\tau_{GW} \approx \tau_{KH}$. In the high mass range $M_{hl}<M_2<M_\star$, when the secondary has constant opacity and radiative structure, the mass-radius exponent is $\varepsilon=4/15$. For lower masses $M_2<M_{hl}$, when the $H^-$ opacity dominates and the surface temperature is roughly constant, $\varepsilon \approx 13/21$. The various evolutionary tracks are illustrated in figure \ref{fig:MassRadius_evolution}. The results obtained in section \ref{subsec:MassRadius_EMRI} are somewhat crude, and rely on simple stellar structure models. While the values we quote for $\varepsilon$ in the different regimes should be considered approximate, our model captures qualitatively the possible evolutionary tracks taken by the system, according to the comparison between the gravitational waves and cooling timescales.

For a wide range of SMBH and stellar masses, L2 mass loss occurs in parallel to the L1 mass transfer, when the star initially overfills its Roche lobe. The presence of this non-conservative mass ejection modifies the frequency evolution timescale ($\tau_f \equiv f/\dot{f}$) by up to a few tens of percent. Correspondingly, the equation for estimating the chirp mass has to be modified
(see equation (\ref{eq:ModifiedChirpMass}) as compared to equation (\ref{eq:ChirpMass})).

We briefly address the expected electromagnetic luminosity of these systems. Perhaps the dominant source of luminosity is the accretion of the mass transferred through L1, at a rate $\dot{M}_{L_1}$, onto the black hole. This may result is a luminosity which is some fraction of $\dot{M}_{L_1} c^2$. However, for our Galactic Centre, this accretion rate is smaller than the direct accretion onto the black hole from its surrounding. We therefore do not expect a noticeable electromagnetic signal even if a mass transferring star currently revolves around the black hole.

In addition, a substantial amount of mass may be ejected outward through L2. The fate of L2 mass ejection has been studied by \cite{Lubow_Shu_1979}, and more recently by \cite{PMT1,PMT2}, who simulated the radiative evolution of the L2 ejecta. These works have shown that L2 mass-loss from systems with $q \lesssim 0.06$ remains marginally-bound. When the mass ejecta can cool down efficiently, an excretion disc is formed around the binary system. According to their work, the resulting luminosity of the excretion disc is roughly 
\begin{equation}
    L \approx \dot{M}_{L_2} v_{orb}^2 \,,
\end{equation}
where $v_{orb}$ is the orbital velocity, and $\dot{M}_{L_2}$ is the mass loss rate through L2.

The expression for this luminosity can be rewritten for a general external driving mechanism with timescale $\tau$ as
\begin{equation}
    L \approx \alpha \frac{M_2 c^2}{\tau} \beta^2 q^{-2/3} \,,
\end{equation}
and for evolution through GW emission (equation \ref{eq:GW_timescale_beta}), the luminosity is given by
\begin{equation}
    L \approx \alpha \frac{M_2 c^2}{\tau_{dyn}} \beta^7 q^{-4/3} \,.
\end{equation}

At the early evolution phase of mass transfer, when the star follows the main-sequence mass-radius relation, we have
\begin{equation}
\frac{L}{L_\odot} \approx 4\times 10^5 \; \alpha \left(\frac{M_2}{M_\odot}\right)^{-1/3} \left(\frac{M_1}{M_{Sgr A*}}\right)^{4/3}
\end{equation}
where as before, $\alpha$ is the fraction of mass lost from L2, compared to the total mass transfer rate from the secondary. Note that \cite{PMT2} have not simulated the ejecta evolution for extreme mass ratio binaries, and expect the tidal torquing to become inefficient as $q \rightarrow 0$. It is therefore not clear whether their results regarding the excretion disc luminosity can be extrapolated to the regimes we study.

A modified version of this theory for L2 mass leakage may prevail for systems with unstable mass transfer as well. This may be relevant to more equal mass binaries, like neutron star mergers. Such a theory may be an important part in describing the dynamical ejecta associated with Macronovas/Kilonovas \citep{Li_Paczynski_1998,Hotokezaka_Piran_2015,Metzger_Kilonova_2016}.

\section*{Acknowledgements}

This research was partially supported by an iCore grant and an ISF grant. We thank Scott Tremaine, Kenta Hotokezaka and Sivan Ginzburg for useful discussions.




\bibliographystyle{mnras}
\bibliography{L2_Leakage_EMRI}



\appendix

\subsection{The synchronization of the secondary} \label{sec:Tidal_Lock}
In this section we examine the assumption that the secondary remains tidally locked as it is transferring mass. As long as sufficient tidal dissipation exists in the secondary, the timescale for tidal locking will be shorter than the orbital evolution timescale. This would imply that as the orbit evolves, the secondary remains synchronized to the changing orbital period. The tidal torque exerted on the secondary at the Roche limit is given by:

\begin{equation}
    \dot{J} \approx -\frac{GM_2^2}{R_2} \theta_{lag}
\end{equation}
where \(\theta_{lag}\) is the tidal lag angle between the elongated axis of the secondary and the line connecting the components' centres of mass. The spin of the secondary is approximately:

\begin{equation}
    J_{spin,2} \approx M_2 R_2^2 \Omega_2
\end{equation}
where \(\Omega_2\) is the secondary's sidereal frequency, which must be smaller than the breakup frequency \(\Omega_2 < \sqrt{G\rho_2}\). Therefore, the synchronization timescale is shorter than
\begin{equation}
    \tau_{syn} < \frac{1}{\sqrt{G\rho_2} } \theta_{lag}^{-1}
\end{equation}

At the Roche limit, the rotation period is approximately \(\omega \approx \sqrt{G\rho_2}\), and so the system synchronizes over at most \(1/\theta_{lag}\) periods. The lag angle of the tidal bulge depends on the rate of dissipation through viscosity in the secondary. Following the common notation introduced by \cite{Goldreich_1963}, this angle is related to the specific dissipation function, $Q$, which is generally frequency dependent, and $\theta_{lag} \sim Q^{-1}$. Previous studies and observational evidence indicate that $10^5 < Q < 10^7$ in main-sequence stars. The condition for maintaining tidal synchronization is thus
\begin{equation}
    Q < \tau/\tau_{dyn} \,.
\end{equation}

For the specific case of a system evolving through gravitational radiation we have from equation \ref{eq:GW_timescale_beta} the condition
\begin{equation}
    Q < \beta^{-5} q^{2/3} \,,
\end{equation}
where $\beta$ is the secondary's escape velocity, relative to the speed of light. Considering a main-sequence star orbiting a black hole at the Roche limit, we find that the secondary remains synchronized as long as
\begin{equation}
\frac{M_1}{M_{Sgr A*}} < 10^4 \left( \frac{Q}{10^7} \right)^{-3/2} \left( \frac{M_2}{M_\odot} \right)^{0.25} \,,
\end{equation}
and so for all relevant regimes, the secondary maintains synchronization throughout the evolution.

\subsection{Comparing the scale height to the L1-L2 gap} \label{appendix:ScaleHeight}

In section \ref{sec:L2 Mass Loss} we derived the conditions for L2 leakage for a secondary with a polytropic structure. We assumed a sharp boundary for the secondary, and found that when the system evolves sufficiently fast, the secondary's surface extends all the way to L2. In this appendix we study the L2 leakage due to an isothermal atmosphere, in a mass transferring system. If the atmospheric scale height is sufficiently large, leakage from L2 occurs even for slow orbital evolution. While this could potentially serve as an alternative channel of L2 leakage, we show that it is unlikely to occur for reasonable surface temperatures.

For a secondary with an isothermal atmosphere of temperature \(T\), we compare the atmospheric scale height, \(H_p\) to \(\Delta R_{L_1L_2}\) - the volumetric radius difference between the L1 and L2 equipotential surfaces.

\begin{equation} \label{eq:ScaleHeight}
    H_p \approx \frac{K_BT}{\mu g}
\end{equation}

where \(K_B\) is the Boltzmann constant, \(\mu\) is the mean atmospheric particle mass and \(g\) is the surface gravity.

\begin{equation}
    \frac{H_p}{R_2} \approx \left(\frac{v_T}{v_{esc}}\right)^2
\end{equation}
where \(v_T\) is the thermal velocity.

Using equation \ref{eq:DeltaRL1RL2}, we write:

\begin{equation}
    \frac{H_p}{\Delta R_{L_1L_2}} \approx \left(\frac{v_T}{v_{esc}}\right)^2 q^{-1/3}
\end{equation}

We compare the above ratio to unity, to determine the temperature at which the atmosphere reaches the L2 surface before decaying significantly. By defining \(\beta_T \equiv v_T/c\) we have:

\begin{equation}
    \frac{H_p}{\Delta R_{L_1L_2}} > 1
\end{equation}

\begin{equation} \label{eq:q_thermal_leakage}
q < (\beta_T/\beta)^6
\end{equation}

For a solar mass main-sequence star, \(T\approx 6000 K\), and so \(\beta_T \approx 2\times10^{-5}\), while \(\beta\approx 1.5\times 10^{-3}\). This implies that a stellar atmosphere could bridge over the gap between the equipotential surfaces of L1 and L2 only for systems with a mass ratio of \(q < 10^{-11}\). Since the mass of the heaviest known SMBHs are of order \(10^{10} M_{\odot}\), it is not likely that L2 mass loss by extended atmospheres will occur in most systems of a SMBH and a main-sequence star.


\bsp	
\label{lastpage}
\end{document}